\documentclass[lettersize,journal]{IEEEtran}
\usepackage{amsmath,amsfonts}
\usepackage{algorithm}
\usepackage{algorithmic}
\usepackage{array}
\usepackage[caption=false,font=normalsize,labelfont=sf,textfont=sf]{subfig}
\usepackage{textcomp}
\usepackage{stfloats}
\usepackage{url}
\usepackage{verbatim}
\usepackage{graphicx}
\usepackage{multirow}
\usepackage{booktabs}
\usepackage{caption}
\usepackage{balance}
\usepackage{tabularx}
\usepackage{cite}
\usepackage{amsthm}
\newtheorem{proposition}{Proposition}[section]
\newtheorem{lemma}{Lemma}[section]
\newtheorem{corollary}{Corollary}[section]

\usepackage[table]{xcolor}
\usepackage{amssymb}
\begin{document}

% \title{STA: A Split Temporal Adversary Framework for Evaluating Vulnerabilities in MultiModal RL Policies}
% \title{STA: A Split Temporal Adversary Method for Evaluating MultiModal Vulnerabilities in RL Policies}
% \title{STA: A Split Temporal Adversary Method for Temporal Vulnerability Evaluation of Multimodal RL Driving Policies}
% \title{A Temporal-Modality-Splitting Adversary Method for Evaluating Vulnerability in Multimodal Reinforcement Learning Driving Policies

% -------------

\title{STA: A Split Temporal Adversary Method for Evaluating Robustness of Multimodal RL-based Driving Policies

% -----------------

% Evaluating Effectiveness/Robustness of Multimodal RL-based Driving Policies: A Split Temporal Adversary Method
}

\author{
Shenghao Xiong, Xiaolin Chang,\thanks{Shenghao Xiong and Xiaolin Chang are with the School of Cyberspace Science and Technology, Beijing Jiaotong University, Beijing, China (e-mails: {25115273, xlchang}@bjtu.edu.cn).}
}

\maketitle
\begin{abstract}
Multimodal reinforcement learning (MRL) offers a promising paradigm for autonomous driving by jointly leveraging complementary sensor observations to learn effective decision-making policies. However, its effectiveness critically depends on temporal alignment across modalities, making temporal inconsistency a fundamental vulnerability.
Even subtle misalignment or latency can corrupt fused representations and ultimately lead to decision failures.
Existing studies have explored adversarial attacks to induce cross-modal temporal inconsistencies, thereby enabling systematic evaluation of the temporal robustness of MRL policies.
However, existing methods remain limited in generality, realism, and adaptivity.
To address these limitations, we propose the Split Temporal Adversary (STA), which formulates multimodal temporal attacks as a unified decision process that jointly optimizes attack timing and modality selection under realistic policy-aware and budget-constrained settings.
STA includes two mechanisms: 
a) a unified multimodal temporal perturbation mechanism that delays observations via historical states, removing the need for modality-specific noise; 
b) a split multimodal adversarial decision mechanism that jointly optimizes attack timing and modality selection through adaptive optimization while leveraging the victim policy's clean actions to identify sensitive decision patterns.
We provide a theoretical analysis of the induced distribution shift and a modality selection principle. 
Experiments in the CARLA simulator show that STA more than doubles the collision rate of the strongest baseline against PPO while using 17.8\% fewer attack attempts, and that it generalizes to a VLM-based driving policy. 
These results reveal temporal consistency as a critical vulnerability, highlighting temporal robustness as essential for safe autonomous driving.
\end{abstract}

\begin{IEEEkeywords}
 Adversarial attacks, autonomous driving, multimodal reinforcement learning, temporal consistency.
\end{IEEEkeywords}

\section{Introduction}
\IEEEPARstart{T}he integration of multimodal perception with reinforcement learning (RL) has emerged as a promising paradigm for developing autonomous driving policies, as it enables vehicles to learn complex behaviors directly from high-dimensional sensor inputs~\cite{VLM-RL}. However, its effectiveness depends on maintaining temporal consistency across heterogeneous sensor inputs. This dependence inherently establishes temporal inconsistency itself as a core vulnerability, since any misalignment,
latency jitter, or asynchrony among sensor modalities can directly corrupt the fused representations and impair decision-making reliability in multimodal decision-making systems. Therefore, rigorous temporal robustness assessment is an essential prerequisite for ensuring the operational safety and reliability of multimodal driving policies in safety-critical deployments~\cite{Qian2025Survey, Jalalipour2026MultimodalRL}. 

Prior work has investigated adversarial attacks to expose vulnerability in multimodal driving policies~\cite{Barach2024multimodal, cheng2024fusion}. 
However, existing approaches exhibit \textbf{three fundamental limitations} that prevent them from effectively exposing this temporal consistency vulnerability. First, most attacks rely on modality-specific perturbations (e.g., image noise or LiDAR point perturbations~\cite{goodfellow2014explaining, kurakin2018adversarial, huang2017adversarial, sun2020stealthy, li2023ats}), which are sensor-specific, lack a unified formulation, and inadequately assess temporal consistency vulnerability. 
Second, they overlook policy awareness and attack budgets, ignoring victim policy feedback and yielding unrealistic threat models~\cite{he2022robust, he2022robustlane, sharif2022adversarial, cai2024adversarial, kulkarni2024crash}. 
Third, temporal perturbation methods rely on random modality selection or fixed injection schedules~\cite{shahriar2025temporal, xiahou2021robust, li2021chronos}, lacking adaptive optimization over attack timing and modality selection.

We identify \textbf{two challenges} in addressing these limitations:

\begin{enumerate}
\item \textbf{Challenge of modality universality.} 
Heterogeneous sensors (e.g., cameras, LiDAR, CAN) exhibit distinct representations and temporal dynamics, creating a conflict between modality-specific semantic validity and cross-modal temporal consistency~\cite{Feng2020Challenge,Liang2026SingleToMulti}. This hinders the design of a unified perturbation across modalities (addressing \textbf{limitation 1}).

\item \textbf{Challenge of adaptive multimodal adversarial decision-making.}
The effectiveness of temporal perturbations depends heavily on both the driving context~\cite{Fan2025Less} and the targeted sensor modality. Under sparse collision-based rewards and finite attack budgets, an adversary must jointly determine \emph{when} to launch an attack and \emph{which} modality to perturb, while estimating policy vulnerability solely from observable interactions without access to internal policy states~\cite{He2024Trustworthy}. This coupling of attack timing, modality selection, and realistic adversarial constraints makes adaptive attack policy optimization particularly challenging (addressing \textbf{limitations 2 and 3}).
\end{enumerate}

To address the above limitations and challenges, we propose the Split Temporal Adversarial (STA) method. \textbf{To the best of our knowledge}, STA is the first adversarial framework that enables unified temporal perturbation across heterogeneous sensor modalities while adaptively optimizing attack timing and modality selection under realistic attack constraints. STA consists of two core mechanisms:

\begin{itemize}
\item \textbf{A unified multimodal temporal perturbation mechanism} that maintains a history buffer and replaces the current observation of a selected modality with its $n$-step delayed counterpart, enabling modality-agnostic temporal attacks across cameras, LiDAR, and CAN signals (addressing \textbf{Challenge 1}).

\item \textbf{A split multimodal adversarial decision mechanism} that decouples attack triggering and modality selection, enabling adaptive optimization of both when to attack and which modality to perturb under a gated learning scheme (addressing \textbf{Challenge 2}).
\end{itemize}

We further analyze the distribution shift induced by temporal inconsistency, bound its propagation through a Lipschitz policy, and characterize modality sensitivity through the policy Jacobian. These results motivate adaptive selection of vulnerable modalities.

Experiments in the CARLA simulator~\cite{carla} show that STA achieves higher attack effectiveness with fewer attempts than the baseline methods and generalizes to a VLM-based driving policy. These results reveal that disruption of synchronized alignment across modalities is a critical vulnerability.

The remainder of this paper is organized as follows. Section II reviews related work. Section III presents the preliminary concepts, Section IV presents the STA method, Section V
describes the theoretical analysis, Section VI reports experimental results, and Section VII concludes the paper and discusses future directions.

\section{Related Work}

Our work lies at the intersection of adversarial attacks on deep reinforcement learning, robustness evaluation of multimodal autonomous driving systems, and temporal attacks in cyber-physical systems.

\subsection{Adversarial Attacks on Deep Reinforcement Learning}

Deep reinforcement learning (DRL) policies are highly vulnerable to adversarial perturbations. Early work~\cite{huang2017adversarial} demonstrated that small observation perturbations can significantly degrade policy performance in Atari environments. Subsequent studies improved attack effectiveness through policy-sensitive objectives~\cite{pattanaik2017robust, sun2020stealthy, li2023ats}, e.g., boosting success rates by 18.7\% via policy-sensitive targeting~\cite{li2023ats}. More advanced formulations model attacker–victim interactions as sequential decision-making processes~\cite{mo2022attacking, domico2025adversarial, Behzadan2019AdversarialRL}.

These DRL attack methods operate in a step-wise manner, perturbing observations independently at each timestep. This design largely ignores long-term temporal dependencies and assumes unrestricted access to all modalities, limiting their applicability to multimodal driving scenarios. A growing body of work has begun to address these limitations by considering temporally correlated perturbations, yet such efforts remain largely confined to single-modality settings~\cite{Yang2020Enhanced, Zhao2020Black}.

\subsection{Robustness of Multimodal Driving Policies}

Modern autonomous driving systems rely heavily on multimodal sensor fusion, introducing new attack surfaces across perception and decision-making pipelines. Prior work has shown that perturbations in one modality can propagate through fusion networks and significantly degrade downstream policy performance~\cite{Barach2024multimodal, cheng2024fusion}.

At the decision-making level, several works formulate adversarial driving as constrained optimization or reinforcement learning problems to induce unsafe behaviors~\cite{he2022robust, sharif2022adversarial, cai2024adversarial, kulkarni2024crash}. In parallel, robustness evaluation and certification methods have been proposed to quantify safety under adversarial conditions~\cite{huang2025commit, wu2025certified, li2025towards}.

However, these studies primarily focus on spatial or scenario-level perturbations, and do not explicitly consider temporal inconsistency across heterogeneous modalities.

\subsection{Temporal Attacks and Time-Delay Vulnerability}

Temporal attacks have proven effective in cyber-physical systems, where injected delays destabilize control loops in power grids, sensor networks, and UAV formations~\cite{xiahou2021robust, zhai2022etd, zhai2023hotd, Song2025TimeDelay}. Translating this threat to autonomous driving, Chronos~\cite{li2021chronos} first demonstrated that software-induced timing delays can degrade perception and planning. Subsequent work has explored broader forms of temporal attacks, including sensor spoofing with time shifts and delayed actuation commands. More broadly, recent studies on optimal attack scheduling in cyber-physical systems have shown that strategic timing can achieve greater disruption with fewer interventions, providing theoretical grounding for adaptive scheduling~\cite{Song2022OptimalTiming}.

However, existing temporal attacks on autonomous driving suffer from two persistent limitations. First, most are evaluated only at the perception level: DejaVu~\cite{shahriar2025temporal} showed that temporal misalignment between camera and LiDAR streams can reduce detection mAP by 88.5\% with a single-frame delay, yet does not assess downstream impact on driving decisions; SlowPerception~\cite{ma2024slowperception} achieved a 97\% collision rate, but its delay schedule is pre-determined rather than adaptively optimized. Second, existing approaches operate in an open-loop manner, with fixed or heuristic-driven delays that lack policy awareness and attack budget constraints.

Consequently, while these works confirm temporal inconsistency as a severe vulnerability, none addresses \emph{adaptive, policy-aware, budget-constrained} temporal attack optimization that jointly selects \emph{when} to attack and \emph{which modality} to delay which our STA method directly fills.

\section{Preliminary}
\label{sec:prelim}
This section first discusses temporal inconsistency vulnerability in multimodal perception systems in Section~\ref{sec:prelim-temporal}. Then, Section~\ref{sec:prelim-pomdp} presents the formulation of the adversarial learning problem as a single-agent POMDP.

\subsection{Temporal Inconsistency Vulnerability}
\label{sec:prelim-temporal}

Temporal consistency is a fundamental assumption in multimodal driving systems: observations from different sensor modalities are assumed to correspond, at the same timestamp, to a unified underlying environmental state. To achieve this, each modality typically maintains a finite FIFO buffer \(Q_m\) that stores the most recent messages, and a synchronization mechanism at the fusion layer matches cross-modal messages based on their timestamps. For a pair of messages from modalities \(m\) and \(l\), a match is accepted if
\begin{equation}
    | t_m^{(i)} - t_l^{(j)} | \leq \tau,
\end{equation}
where \(t_m^{(i)}\) denotes the timestamp of the \(i\)-th message from modality \(m\), and \(\tau\) is a pre-defined tolerance threshold. The timestamp difference for accepted pairs is bounded by \(\tau\), which we refer to as the \emph{reported relative temporal mismatch} (R-RTM).

However, this alignment mechanism relies solely on timestamp headers and does not verify the semantic content of the observations. As a result, it only guarantees that the \emph{reported} timestamps are within tolerance, not that the observations correspond to the same physical instant. This creates a vulnerability: if an adversary can manipulate the timestamps or delay the transmission of messages, the synchronizer may be induced to pair observations from different time steps, resulting in \emph{semantic relative temporal mismatch} (S-RTM):
\begin{equation}
    \Delta_{\text{sem}}^{(i,j)} = | u_m^{(i)} - u_l^{(j)} |,
\end{equation}
where \(u_m^{(i)}\) denotes the true acquisition time of the measurement. Under adversarial interference, \(\Delta_{\text{sem}}\) can be arbitrarily large even when the R-RTM remains within \(\tau\).

We refer to the deviation between \(\Delta_{\text{sem}}\) and \(\Delta_{\text{rep}}\) as \emph{temporal inconsistency}. When such inconsistency occurs across modalities, the fused observation no longer represents a single coherent environment state, potentially leading to erroneous downstream decisions.

\subsection{Problem Formulation as a POMDP}
\label{sec:prelim-pomdp}

We model the adversary--victim interaction as a POMDP
\begin{equation}
\mathcal{M} = \langle \mathcal{S}, \mathcal{O}, \mathcal{A}, \mathcal{T}, \mathcal{R}, \gamma \rangle,
\end{equation}
where the victim is part of the environment; $\mathcal{S}$, $\mathcal{O}$, and $\mathcal{A}$ are the latent state, adversarial observation, and action spaces; $\mathcal{T}$ and $\mathcal{R}$ are the transition and reward functions; and $\gamma$ is the discount factor. The adversary learns $\pi:\mathcal{O}\to\Delta(\mathcal{A})$ to maximize
\begin{equation}
J(\pi) = \mathbb{E}_{\pi} \left[ \sum_{t=0}^{T-1} \gamma^t r_t \right].
\end{equation}
Section~\ref{sec:method} instantiates these elements with policy-aware observations, split attack actions, and a finite budget.

\section{Split Temporal Adversary Method}
\label{sec:method}

This section first overviews the proposed STA. Then, the two core mechanisms of STA are detailed.

\subsection{Overview}

We evaluate the temporal robustness of multimodal RL driving policies against adversarial temporal inconsistency, where an adversary selectively disrupts cross-modal alignment under a limited attack budget. We assume: (i) the driving policy \(\pi_{\text{driving}}\) is fixed during evaluation~\cite{pattanaik2017robust, huang2017adversarial}; (ii) temporal delays are bounded within horizon \(n\)~\cite{li2021chronos, shahriar2025temporal}; and (iii) all sensors operate at a fixed synchronized rate~\cite{Barach2024multimodal, cheng2024fusion}. The adversary has a finite budget \(B\) and aims to induce failures (e.g., collisions).

Split Temporal Adversary (STA) consists of two coordinated mechanisms:
\begin{enumerate}
    \item \textbf{Unified Multimodal Temporal Perturbation Mechanism}: maintains a multimodal history buffer and replaces the current observation of a selected modality with its delayed historical counterpart.

     \item \textbf{Split Multimodal Adversarial Decision Mechanism}: jointly optimizes attack triggering and modality selection through a split decision architecture, while incorporating policy-aware state information and attack budget constraints to learn adaptive attack strategies under sparse rewards.
\end{enumerate}
Together, these two mechanisms enable STA to generate adaptive, modality-agnostic temporal attacks under realistic adversarial constraints. We describe each mechanism in detail in the following subsections.

\IfFileExists{picture/framework0714.png}{%
\begin{figure*}[t]
\centering
\includegraphics[width=1\textwidth]{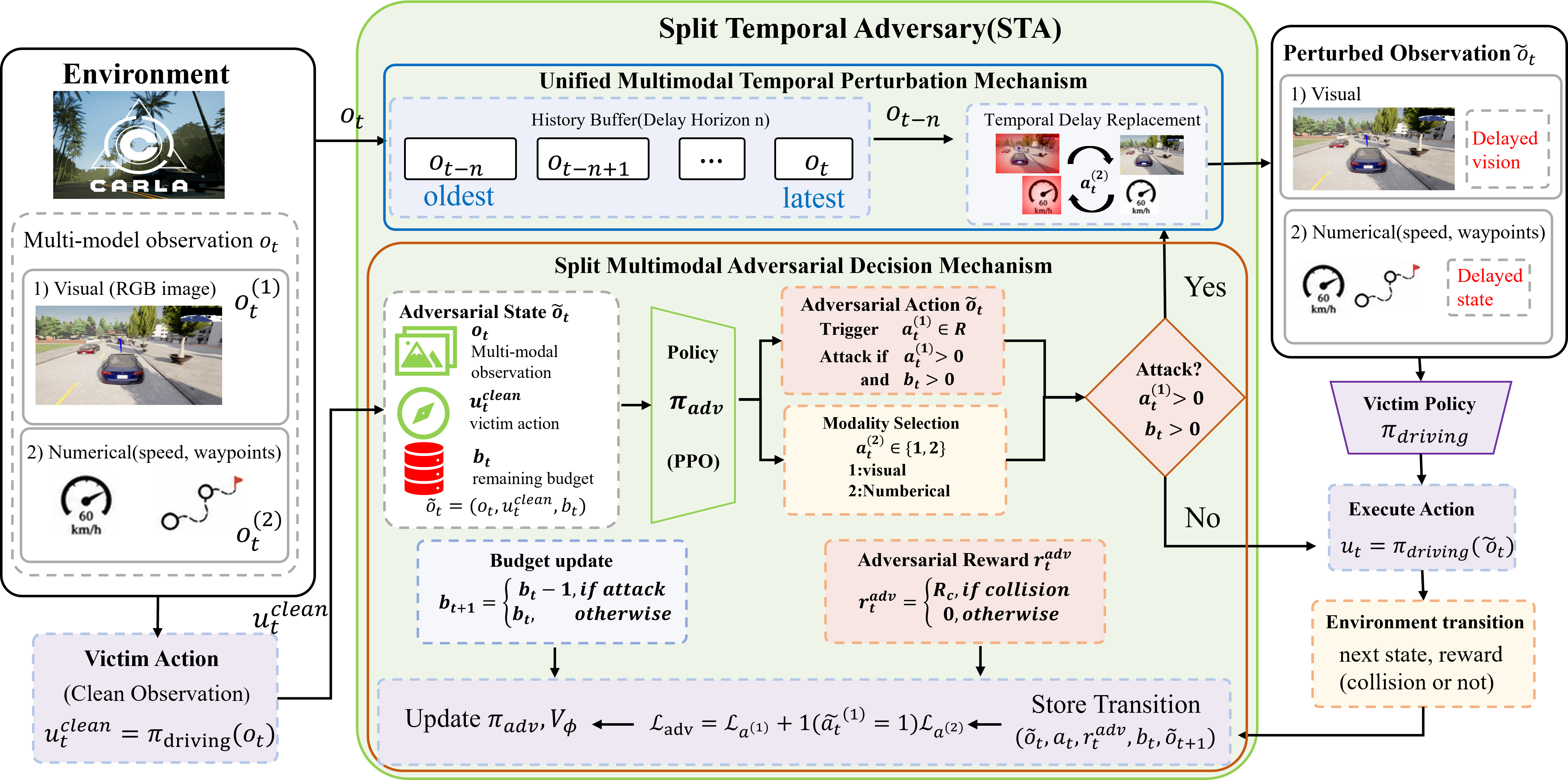}
\caption{Overview of the proposed Split Temporal Adversary (STA). At each time step, the adversary observes the current state and driving policy information, then decides whether to attack based on the attack budget and policy awareness. If activated, it selects which modality to manipulate (visual or numerical) and replaces the selected modality with a delayed observation from the history buffer. This creates cross-modal temporal misalignment that leads to erroneous control decisions by the driving policy.}
\label{fig:algorithm_framework}
\end{figure*}
}{}

\subsection{Unified Multimodal Temporal Perturbation Mechanism}
\label{subsec:multimodal}
This mechanism aims to operate uniformly across heterogeneous sensor inputs by treating temporal inconsistency as a universal attack surface in multimodal reinforcement learning, instead of designing modality-specific noise functions. For instance, for a selected modality at time step \(t\), the adversary replaces its current observation \(o_t^m\) with a historical observation \(o_{t-n}^m\) retrieved from a fixed-length buffer of past states. This operation preserves intra-modality semantic consistency while introducing cross-modal temporal misalignment at the fusion level.

\subsubsection{Multimodal Observation Space}

The driving agent perceives the environment through a multimodal observation space composed of visual and numerical inputs. At each discrete time step \(t\), the clean observation is defined as:
\begin{equation}
o_t = \left( o_t^{(v)}, o_t^{(n)} \right),
\end{equation}
where \(o_t^{(v)}\) denotes the visual observation from a front-view RGB camera, and \(o_t^{(n)}\) represents a low-dimensional numerical state vector including vehicle dynamics and planned waypoints.

To obtain a compact representation, the driving policy encodes each modality into latent embeddings:
\begin{equation}
z_t^{(v)} = f_{\mathrm{vis}}(o_t^{(v)}), \quad
z_t^{(n)} = f_{\mathrm{num}}(o_t^{(n)}),
\end{equation}
which are concatenated into a unified representation:
\begin{equation}
h_t = \left[ z_t^{(v)}; z_t^{(n)} \right].
\end{equation}

This early-fusion design directly operates on concatenated latent features for decision-making, without explicitly modeling cross-modal temporal alignment, which motivates the study of temporal inconsistency in multimodal reinforcement learning policies.

\subsubsection{History Buffer}

The adversary maintains a circular history buffer \(\mathcal{H}_t\) that stores the most recent \(n\) multimodal observations. Each entry corresponds to a complete observation tuple:
\begin{equation}
\mathbf{o}_t = \left( o_t^{(v)}, o_t^{(n)} \right),
\end{equation}
and the buffer is defined as:
\begin{equation}
\mathcal{H}_t = \{\mathbf{o}_{t-n+1}, \dots, \mathbf{o}_t\}.
\end{equation}

This design maintains a shared temporal repository for all modalities, enabling the adversary to sample consistent historical states without maintaining separate modality-specific buffers.

When a temporal attack is triggered on modality \(m\), the adversary samples a past observation \(\mathbf{o}_{t-k} \in \mathcal{H}_t\) and extracts the corresponding modality component:
\begin{equation}
o_{t-k}^{(m)} = \mathrm{Proj}_m(\mathbf{o}_{t-k}),
\end{equation}
where \(\mathrm{Proj}_m(\cdot)\) denotes modality projection.

The extracted delayed observation \(o_{t-k}^{(m)}\) is then used in the temporal replacement operation (see Eq.~\eqref{eq:adv_delay}). This formulation enables modality-agnostic temporal sampling while preserving cross-modal alignment within each historical snapshot.

\subsubsection{Unified Delay Perturbation}

Given an attack decision at time step \(t\), STA applies a temporal replacement operator to the selected modality \(m\). The perturbed observation is defined as:
\begin{equation}
\tilde{o}_t^{(m)} = 
\begin{cases}
o_{t-n}^{(m)}, & \text{if } m \text{ is selected for attack at } t, \\
o_t^{(m)}, & \text{otherwise},
\end{cases}
\label{eq:adv_delay}
\end{equation}
and the resulting multimodal observation becomes \(\tilde{o}_t = \{\tilde{o}_t^{(m)}\}_{m=1}^M\).

This operator defines a modality-agnostic temporal perturbation mechanism that replaces the current observation with a historical counterpart without introducing additive noise or modality-specific transformations. As a result, the attack effect is induced solely through cross-modal temporal misalignment at the fusion level.

\subsubsection{Semantic Plausibility vs. Temporal Inconsistency}

Each delayed observation \(o_{t-n}^{(m)}\) is a valid sensor reading from a past time step and therefore remains semantically meaningful in isolation. However, the resulting multimodal observation \(\tilde{o}_t\) combines signals originating from different time steps, leading to a violation of temporal consistency assumptions in multimodal fusion.

Unlike additive perturbations that alter the content of individual modalities, the proposed delay-based attack preserves intra-modality semantics while introducing cross-modal temporal misalignment, which can destabilize decision-making in temporally sensitive driving policies.

\subsubsection{Fixed Temporal Delay Mechanism}

We adopt a fixed temporal delay of \(n\) steps as the core perturbation strategy, where the selected modality is replaced with its corresponding historical observation from \(n\) time steps earlier.

Formally, this corresponds to selecting the oldest accessible entry within the history buffer, ensuring a consistent and controllable temporal shift across all attack instances.

This deterministic delay mechanism provides a stable and reproducible perturbation strength, and its impact on policy behavior is further analyzed in Section~\ref{sec:theoretical_analysis}.

\subsection{Split Multimodal Adversarial Decision Mechanism}
\label{sec:split_decision}

Based on the POMDP formulation introduced in Section~III, this mechanism instantiates the adversarial decision process for multimodal temporal attacks by augmenting the observation space with policy-aware information and redesigning the action space into a split decision architecture. The resulting adversarial policy jointly determines when to launch an attack and which modality to perturb while operating under a finite attack budget.

\subsubsection{Policy-Aware Observation}

The adversarial policy conditions its decisions on an augmented observation constructed from the multimodal observation defined in Section~\ref{subsec:multimodal}. The adversarial observation is defined as

\begin{equation}
\hat{o}_t=(o_t,u_t^{\mathrm{clean}},b_t),
\end{equation}

where \(o_t\) denotes the multimodal observation, \(u_t^{\mathrm{clean}}=\pi_{\mathrm{driving}}(o_t)\) is the clean action generated by the victim driving policy, and \(b_t\) denotes the remaining attack budget.

The clean action provides an estimate of the victim policy's local decision tendency, while the remaining budget enables long-horizon planning under realistic resource constraints.

\subsubsection{Split Hierarchical Action Space}

Given the policy-aware observation, the adversarial policy outputs a split decision consisting of an attack trigger and a modality selector:

\begin{equation}
a_t=(a_t^{\mathrm{trig}},a_t^{\mathrm{mod}}),
\end{equation}

where \(a_t^{\mathrm{trig}}\in\mathbb{R}\) determines whether an attack is executed, and \(a_t^{\mathrm{mod}}\in\{1,2\}\) specifies the target modality.

The effective attack trigger is obtained as

\begin{equation}
\tilde{a}_t^{\mathrm{trig}}
=
\mathbf{1}
\left(
a_t^{\mathrm{trig}}>0
\land
b_t>0
\right),
\end{equation}

where \(b_t\) is the remaining attack budget. When
\(\tilde{a}_t^{\mathrm{trig}}=1\),
the selected modality
\(a_t^{\mathrm{mod}}\)
is perturbed using the temporal replacement mechanism described in section \ref{subsec:multimodal}.

This factorization decouples attack triggering from modality selection, allowing each decision to be optimized independently while jointly forming a complete adversarial strategy.

\subsubsection{Budget-Aware Optimization}

The remaining attack budget evolves according to

\begin{equation}
b_{t+1}
=
\begin{cases}
b_t-1,
&
\text{if }
\tilde{a}_t^{\mathrm{trig}}=1,
\\
b_t,
&
\text{otherwise}.
\end{cases}
\end{equation}

The adversary receives a sparse collision-based reward,

\begin{equation}
\label{eq:reward}
r_t^{\mathrm{adv}}
=
\begin{cases}
R_c,
&
\text{if a collision occurs at }t,
\\
0,
&
\text{otherwise},
\end{cases}
\end{equation}

where a collision occurring within a horizon of \(T\) steps after an attack is regarded as a successful attack and receives reward \(R_c\).

The adversarial policy aims to maximize the expected discounted return

\begin{equation}
J(\pi_{\mathrm{adv}})
=
\mathbb{E}_{\pi_{\mathrm{adv}}}
\left[
\sum_{t=0}^{T-1}
\gamma^t
r_t^{\mathrm{adv}}
\right].
\end{equation}

To address the credit assignment problem introduced by the split action space, we adopt a gated optimization objective,

\begin{equation}
\label{eq:loss}
\mathcal{L}_{\mathrm{adv}}
=
\mathcal{L}_{\mathrm{trig}}
+
\mathbf{1}
(\tilde{a}_t^{\mathrm{trig}}=1)
\mathcal{L}_{\mathrm{mod}},
\end{equation}

where
\(\mathcal{L}_{\mathrm{trig}}\)
and
\(\mathcal{L}_{\mathrm{mod}}\)
are the surrogate losses for the trigger and modality branches, respectively. The gating indicator ensures that the modality selector is updated only when an attack is executed, allowing optimization to focus on decision-critical interactions.

\subsubsection{Policy Optimization}

The adversarial policy is optimized using Proximal Policy Optimization (PPO)~\cite{schulman2017proximal}. During training, trajectories are collected by interacting with the victim driving policy under the attack budget constraint, and the gated objective in Eq.~(\ref{eq:loss}) is optimized using the standard PPO clipped surrogate objective for both decision branches.

\subsection{Overall Workflow of Split Temporal Adversary (STA)}

Algorithm~\ref{alg:sta_training} summarizes the overall workflow of STA by integrating the two proposed mechanisms. At each interaction step, the unified multimodal temporal perturbation mechanism maintains the history buffer and performs temporal replacement on the selected modality, while the split multimodal adversarial decision mechanism constructs the policy-aware adversarial state and determines both attack triggering and modality selection under budget constraints. The adversarial policy is optimized using PPO, whereas the victim driving policy remains fixed throughout training.

\begin{algorithm}[t]
\caption{Training Procedure of Split Temporal Adversary (STA)}
\label{alg:sta_training}
\begin{algorithmic}[1]
\STATE \textbf{Input:} Driving policy $\pi_{\text{driving}}$, delay horizon $n$, budget $B$
\STATE Initialize adversary policy $\pi_{\text{adv}}$ and value network $V_{\phi}$
\FOR{each episode}
    \STATE Reset environment; initialize history buffer $\mathcal{H}$
    \STATE Set budget $b \leftarrow B$
    \FOR{$t = 1$ to $T_{\max}$}
        \STATE Observe $o_t$ from environment
        \STATE Compute $u_t^{\text{clean}} = \pi_{\text{driving}}(o_t)$
        \STATE Form adversarial state $\hat{o}_t = (o_t, u_t^{\text{clean}}, b)$
        \STATE Sample action $a_t = (a_t^{\text{trig}}, a_t^{\text{mod}}) \sim \pi_{\text{adv}}(\cdot|\hat{o}_t)$
        \STATE Initialize $\tilde{o}_t \leftarrow o_t$

        \IF{$a_t^{\text{trig}} > 0$ \AND $b > 0$}
            \STATE Retrieve $o_{t-n}^{(a_t^{\text{mod}})}$ from $\mathcal{H}$
            \STATE Replace $\tilde{o}_t^{(a_t^{\text{mod}})} \leftarrow o_{t-n}^{(a_t^{\text{mod}})}$
            \STATE $b \leftarrow b - 1$
        \ENDIF

        \STATE Execute action $u_t = \pi_{\text{driving}}(\tilde{o}_t)$ in environment
        \STATE Receive reward $r_t^{\text{adv}}$ and next observation $o_{t+1}$
        \STATE Store transition $(\hat{o}_t, a_t, r_t^{\text{adv}}, \hat{o}_{t+1})$
        \STATE Update $\mathcal{H} \leftarrow \mathcal{H} \cup \{o_t\}$ (FIFO)
    \ENDFOR

    \STATE Compute advantages using GAE
    \STATE Update $\pi_{\text{adv}}, V_{\phi}$ via PPO objective
\ENDFOR
\end{algorithmic}
\end{algorithm}

Following Fan et al.~\cite{Fan2025Less}, we further adopt a trajectory cropping strategy to improve the training efficiency of the split adversarial policy under finite attack budgets. Once the budget \(b_t\) is exhausted, the adversary can no longer execute further attacks, and the remaining interaction may not provide meaningful learning signals for the adversarial policy. To improve training efficiency, we truncate the episode at the point of budget exhaustion. If the final executed attack results in a collision, the episode is terminated immediately and the trajectory is stored for policy optimization. Otherwise, the environment is allowed to continue for up to \(T\) additional steps to capture delayed effects of previous attacks, and any collision occurring within this window is still considered as a valid adversarial success signal. This design ensures that STA focuses on learning effective attack timing under strict budget constraints while avoiding redundant trajectory collection after budget depletion, and is particularly important for PPO stability under sparse reward settings.

\section{Theoretical Analysis}
\label{sec:theoretical_analysis}

Consider a multimodal system with $M$ sensors. At time $t$, the state $s_t$ generates modality‑specific observations $o_t^{(m)} \sim \mathcal{O}^{(m)}(s_t)$. The policy $\pi$ maps the synchronous tuple $o_t=(o_t^{(1)}\!,\ldots,o_t^{(M)})$ to action $a_t$ and is assumed locally Lipschitz. A temporal adversary delays modality $m$ by $k$ steps, producing $\tilde{o}_t$ where $\tilde{o}_t^{(m)} = o_{t-k}^{(m)}$ and other modalities unchanged. We analyze the attack along three dimensions:
\begin{enumerate}
    \item Distribution shift (Proposition~\ref{prop:distribution_shift});
    \item Perturbation propagation (Proposition~\ref{prop:perturbation_bound});
    \item Modality‑wise sensitivity (Lemma~\ref{lemma:local_linear}, Corollary~\ref{cor:modal_selection}).
\end{enumerate}

\subsection{Distribution Shift Induced by Temporal Inconsistency}

\begin{proposition}[Distribution Shift Lower Bound]
\label{prop:distribution_shift}
Let $\mathcal{O}(s_t)$ be the nominal joint distribution. Under a $k$-step delay on modality $m$, the perturbed distribution is
\begin{equation}
\mathcal{P}_{\tilde{o}_t}^{(m)} = \mathcal{O}^{(m)}(s_{t-k}) \times \mathcal{O}^{(\neg m)}(s_t).
\end{equation}
Then
\begin{equation}
D_{\mathrm{TV}}\!\bigl(\mathcal{P}_{\tilde{o}_t}, \mathcal{O}(s_t)\bigr)
\ge \mathbb{E}_{s_t,s_{t-k}}\!\Bigl[ D_{\mathrm{TV}}\!\bigl(\mathcal{O}^{(m)}(s_{t-k}), \mathcal{O}^{(m)}(s_t)\bigr) \Bigr].
\end{equation}
\end{proposition}

\begin{proof}[\textbf{Proof}]
Only modality $m$ is altered, so the total variation is at least the per‑modality divergence. Restricting the supremum to sets of the form $A = \mathcal{A}^{(m)}\!\times\!\mathcal{O}^{(\neg m)}$ directly yields $D_{\mathrm{TV}}(\mathcal{O}^{(m)}(s_{t-k}),\mathcal{O}^{(m)}(s_t))$; taking expectation over states gives the bound.
\end{proof}

\noindent \textbf{Interpretation.}
Temporal perturbation creates a state-conditioned mismatch in the attacked modality. In rapidly evolving scenarios, $\|s_t-s_{t-k}\|$ can grow, amplifying the shift and degrading policy performance.

\subsection{Perturbation Propagation via Lipschitz Continuity}

\begin{proposition}[Perturbation Propagation Bound]
\label{prop:perturbation_bound}
Assume $\pi$ is $L$-Lipschitz. For the delayed observation $\tilde{o}_t$ (modality $m$ delayed by $k$),
\begin{equation}
d_{\mathcal{A}}(\pi(\tilde{o}_t),\pi(o_t)) \le L\cdot d_{\mathcal{O}}(\tilde{o}_t,o_t).
\end{equation}
If the observation metric is additive, $d_{\mathcal{O}}(o,o')=\sum_i d_{\mathcal{O}}^{(i)}(o^{(i)},o'^{(i)})$, then only the attacked modality contributes:
\begin{equation}
d_{\mathcal{O}}(\tilde{o}_t,o_t) = d_{\mathcal{O}}^{(m)}(o_{t-k}^{(m)},o_t^{(m)}) \triangleq \Delta_m(t,k).
\end{equation}
\end{proposition}

\begin{proof}[\textbf{Proof}]
The Lipschitz condition directly gives the inequality. For the additive metric, $\tilde{o}_t$ and $o_t$ differ only in modality $m$, so all other terms vanish.
\end{proof}

\noindent \textbf{Interpretation.}
Attack impact is governed by policy sensitivity $L$ and the temporal variation $\Delta_m(t,k)$ of the delayed modality. This motivates selecting a delay that creates a substantial temporal discrepancy while retaining a valid historical observation.

\subsection{Differential Impact Across Modalities}

We assume an early‑fusion architecture: modality embeddings $z_t^{\text{vis}}, z_t^{\text{num}}$ are concatenated into $h_t$, and $\pi$ is differentiable.

\begin{lemma}[Local Linear Approximation]
\label{lemma:local_linear}
For a small perturbation $\delta h$,
\begin{equation}
\pi(h_t+\delta h) \approx \pi(h_t) + J(h_t)\,\delta h,
\end{equation}
where $J(h_t)=[\frac{\partial\pi}{\partial z^{\text{vis}}},\frac{\partial\pi}{\partial z^{\text{num}}}]$ is the Jacobian.
\end{lemma}
\begin{proof}[\textbf{Proof}]
Follows from Taylor's theorem with continuity of $J$; the remainder is $o(\|\delta h\|)$.
\end{proof}

\begin{proposition}[Modality‑wise Sensitivity]
\label{prop:fusion_weights}
Attacking modality $m$ yields
\begin{equation}
\|\pi(\tilde{h}_t)-\pi(h_t)\| \approx \|J_m(h_t)(z_{t-k}^{(m)}-z_t^{(m)})\|,
\end{equation}
and in expectation,
\begin{equation}
\mathbb{E}[\|\Delta a_t\|] \le \mathbb{E}\bigl[\|J_m(h_t)\|\cdot\|\Delta z_t^{(m)}\|\bigr], \quad \Delta z_t^{(m)} = z_{t-k}^{(m)}-z_t^{(m)}.
\end{equation}
\end{proposition}
\begin{proof}[\textbf{Proof}]
From Lemma~\ref{lemma:local_linear}, $\pi(\tilde{h}_t)-\pi(h_t) \approx J_m(h_t)\Delta z_t^{(m)}$ (only modality $m$ changes). Taking norms and expectation yields the bound.
\end{proof}

\begin{corollary}[Modality Selection Principle]
\label{cor:modal_selection}
A modality is more influential if it has larger local sensitivity $\|J_m\|$ and larger temporal variation $\mathbb{E}[\|\Delta z_t^{(m)}\|]$. For modalities $i,j$, if $\|J_i\|>\|J_j\|$ and temporal variations are comparable, attacking $i$ induces larger expected action deviation.
\end{corollary}

\noindent \textbf{Interpretation.}
Attack effectiveness depends on both policy sensitivity and intrinsic temporal variation of each modality. In practice, numerical modalities often exhibit higher sensitivity due to direct influence on control signals (e.g., speed), explaining their elevated vulnerability under temporal misalignment.

\section{Experiments}
\label{sec:experiments}

\begin{table*}[t]
\centering
\caption{Main results in Town02 dense traffic. CR: collision rate (\%); ANA: average number of attacks. Mean $\pm$ variance over five seeds.}
\label{tab:main_results}
\renewcommand{\arraystretch}{1.1}
\begin{tabular}{lcccccc}
\toprule
\textbf{Method} 
& \multicolumn{2}{c}{\textbf{PPO}} 
& \multicolumn{2}{c}{\textbf{SAC}} 
& \multicolumn{2}{c}{\textbf{DARRL}} \\
\cmidrule(lr){2-3} \cmidrule(lr){4-5} \cmidrule(lr){6-7}
& CR (\%) & ANA & CR (\%) & ANA & CR (\%) & ANA \\
\midrule

Clean
& 6.8 $\pm$ 0.000 & 0 
& 2.8 $\pm$ 0.000 & 0 
& 2.8 $\pm$ 0.000 & 0 \\

RTA 
& 4.0 $\pm$ 0.001 & 29.524 $\pm$ 0.310
& 4.0 $\pm$ 0.000 & 29.820 $\pm$ 0.052
& 4.4 $\pm$ 0.000 & 29.760 $\pm$ 0.090 \\

RTAM 
& 12.0 $\pm$ 0.001 & 28.864 $\pm$ 0.136
& 5.2 $\pm$ 0.000 & 29.660 $\pm$ 0.050
& 6.8 $\pm$ 0.001 & 29.532 $\pm$ 0.141 \\

ADV 
& 16.8 $\pm$ 0.000 & 29.396 $\pm$ 0.028
& 16.8 $\pm$ 0.002 & 29.532 $\pm$ 0.153
& \textbf{32.2 $\pm$ 0.002} & 28.932 $\pm$ 0.041 \\

\textbf{Ours (STA)} 
& \textbf{35.6 $\pm$ 0.001} & \textbf{24.168 $\pm$ 0.478} 
& \textbf{18.2 $\pm$ 0.001} & \textbf{25.628 $\pm$ 0.136} 
& \textbf{18.6 $\pm$ 0.001} & \textbf{26.124 $\pm$ 0.174} \\

\bottomrule
\end{tabular}
\end{table*}

\begin{table}[t]
\centering
\caption{Results on the VLM-based driving policy in Town02 dense traffic (mean $\pm$ variance over five seeds).}
\label{tab:vlm_results}
\begin{tabular}{lcc}
\toprule
Method & CR (\%) & ANA \\
\midrule
No Attack           & 2.00 $\pm$ 0.00 & -- \\
RTA                 & 0.00 $\pm$ 0.00 & 30.00 $\pm$ 0.31 \\
RTAM                & 2.00 $\pm$ 0.00 & 29.90 $\pm$ 0.14 \\
ADV                 & 2.00 $\pm$ 0.00 & 29.95 $\pm$ 0.03 \\
\textbf{STA (Ours)} & \textbf{62.00 $\pm$ 0.00} & \textbf{26.98 $\pm$ 0.47} \\
\bottomrule
\end{tabular}
\end{table}

We evaluate whether STA (i) induces more safety-critical failures than spatial and random temporal baselines, (ii) uses a limited attack budget more efficiently, and (iii) transfers across victim-policy families.

\subsection{Experimental Setup}
We use a black-box threat model in which the adversary observes the victim's output action but not its action distribution or parameters. Experiments use CARLA 0.9.13, Town02, dense traffic, and fixed clear weather. Victims receive an RGB image plus speed and waypoint states and output continuous steering and throttle. We evaluate PPO (trained for 440,000 steps), SAC (235,000 steps), and our CARLA reimplementation of DARRL~\cite{He2024Trustworthy}. Each victim is trained on clean observations and frozen before adversary training.

STA is trained with PPO for 800,000 steps. Unless stated otherwise, the delay horizon is $n=10$, the history-buffer length is 10, the attack budget is $B=30$, and an episode lasts at most 50 steps. Experiments ran on Intel Xeon Gold 6230 CPUs and NVIDIA RTX 4090 GPUs.

\paragraph{Baselines.}
We compare against the following baselines:
\begin{itemize}
    \item \textbf{Clean:} Clean execution without any intervention.
    \item \textbf{RTA (Random Temporal Attack):} Randomly selects timesteps to apply a fixed $n$-step delay to a random modality~\cite{ma2024slowperception}.
    \item \textbf{RTAM (Random Temporal Attack on all Modalities):} Randomly selects timesteps to apply a fixed $n$-step delay to both numerical and visual modalities~\cite{shahriar2025temporal}.
    \item \textbf{ADV (PGD-based Adversary):} A learned attacker that applies PGD perturbations and adaptively decides when to attack~\cite{domico2025adversarial}.
\end{itemize}

\paragraph{Evaluation Metrics.}
We report collision rate (CR) and average number of executed attacks (ANA). Results aggregate 50 evaluation episodes for each of five independent seeds; the table reports the mean and variance across seeds.

\begin{itemize}
    \item \textbf{Collision Rate (CR):} The percentage of episodes that terminate in a collision. A higher CR indicates a more effective attack at inducing safety-critical failures.
    \item \textbf{Average Number of Attacks (ANA):} The mean number of attacks executed per episode. When combined with a high CR, a lower ANA reflects superior strategic allocation of the limited attack budget.
\end{itemize}

\subsection{Main Results}
Table~\ref{tab:main_results} shows that STA is the strongest attack on PPO and SAC while using fewer interventions than the baselines. The DARRL result also reveals policy-dependent vulnerability: temporal and spatial attacks expose different failure modes.

On PPO, STA raises CR from 16.8\% for the strongest baseline (ADV) to 35.6\%, while reducing ANA from 29.396 to 24.168 (17.8\% fewer interventions). The random temporal baselines nearly exhaust the budget yet remain substantially less effective, supporting the value of policy-aware attack timing and modality selection.

The outcome is policy-dependent. STA is also the strongest attack on SAC, although its advantage over ADV is smaller (18.2\% versus 16.8\%). DARRL instead remains more vulnerable to ADV (32.2\%) than to STA (18.6\%). Thus, robustness to spatial perturbations does not imply robustness to temporal inconsistency, and evaluation should cover both threat models.

\subsection{VLM-Based Driving Strategy Evaluation}

We also evaluate the VLM-based driving policy of~\cite{VLM-RL} under the same Town02 configuration ($B=30$, $n=10$). STA delays image frames; the textual instruction is fixed at episode initialization and is therefore not attacked.

As shown in Table~\ref{tab:vlm_results}, the clean, random-delay, and PGD conditions yield 0--2\% CR, whereas STA reaches 62\% CR with 26.98 attacks per episode. This result suggests that policy-aware temporal disruption can expose failure modes beyond conventional RL policies, although broader evaluation across VLM architectures remains future work.

\section{Conclusion and Future Work}

In this work, we study temporal vulnerability in multimodal RL-based driving policies. Unlike existing works that focus on spatial perturbations, we introduce a threat model that leverages structured temporal misalignment to expose hidden failure modes. We propose STA, a policy-aware and budget-constrained attack framework that selectively delays modality-specific observations, preserving per-modality semantic integrity while inducing cross-modal inconsistency. In CARLA, STA achieves a 35.6\% collision rate on PPO---more than double that of the strongest baseline---while requiring 17.8\% fewer interventions. Its effectiveness on a VLM-based driving policy further suggests that temporal inconsistency is relevant beyond conventional RL. The policy-dependent results also show that resistance to spatial perturbations does not guarantee robustness against temporal inconsistency.

Future work will explore more constrained black-box settings with limited query access, adaptive and learnable delay strategies, and extensions to richer sensor modalities such as LiDAR, radar, and GPS. We also plan to investigate the applicability of STA in real-world and multi-agent driving environments to better understand temporal vulnerability under practical deployment conditions.

\end{document}